\documentclass[epj-spec]{svjour}
\usepackage{graphicx}
\begin{document}
\title{Precise determination of $h/m_{\rm Rb}$ using Bloch oscillations and atomic interferometry: a mean to deduce the fine structure constant}
\author{M. Cadoret\inst{1}\and E. de Mirandes\inst{1} \and P. Clad\'e\inst{1} \and S. Guellati-Kh\'elifa\inst{2} \and C. Schwob\inst{1} \and F. Nez\inst{1} \and L. Julien\inst{1} \and F. Biraben\inst{1}\fnmsep\thanks{\email{biraben@spectro.jussieu.fr}}}
\institute{Laboratoire Kastler Brossel, ENS, CNRS, UPMC, 4 place Jussieu, 75252 Paris CEDEX 05, France\and INM, Conservatoire National des Arts et M\'etiers, 61 rue Landy, 93210 La plaine Saint Denis, France}
\abstract{We use Bloch oscillations to transfer coherently many photon momenta to atoms. Then we can measure accurately the ratio $h/m_{\rm Rb}$ and deduce the fine structure constant $\alpha$. The velocity variation due to the Bloch oscillations is measured thanks to Raman transitions. In a first experiment, two Raman $\pi$ pulses are used to select and measure a very narrow velocity class. This method yields to a value of the fine structure constant $\alpha^{-1}= 137.035\,998\,84\,(91)$ with a relative uncertainty of about 6.6 ppb. More recently we use an atomic interferometer consisting in two pairs of $\pi$/2 pulses. We present here the first results obtained with this method.
} 
\maketitle
\section{Introduction}
\label{intro}
The fine structure constant $\alpha$ is the fundamental physical constant characterizing the strength of the electromagnetic interaction. It is a dimensionless quantity independent of the system of units used. It is defined as:
\begin{equation}
\alpha=\frac{e^2}{4\pi\epsilon_0 \hbar c}
\end{equation}
where $\epsilon_0$ is the permittivity of vacuum, \emph{c} is the
speed of light, \emph{e} is the electron charge and
\emph{$\hbar=h/2\pi$} is the reduced Planck constant. The fine structure constant is a corner stone of the adjustment of the fundamental physical constants \cite{{codata02},{codata06}}. Its value is obtained from experiments in different domains of physics, as the quantum Hall effect and Josephson effect in solid state physics, or the measurement of the muonium hyperfine structure in atomic physics.
\begin{figure}
\centering
\includegraphics[width=8cm]{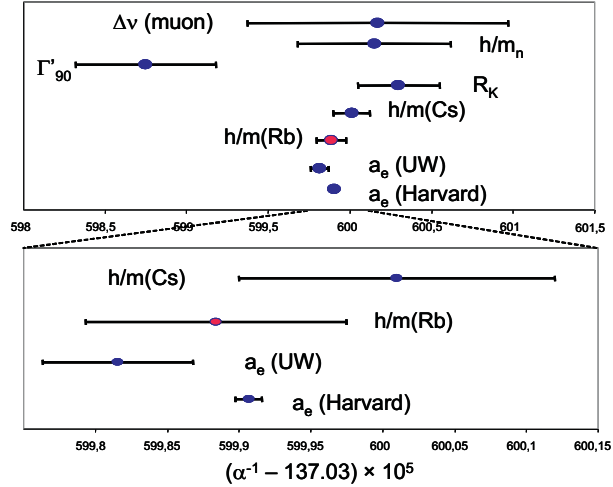}
\caption{Determinations of the fine structure constant in different domains of physics. The most precise measurements used in the CODATA are shown in the lower part of the figure. They are deduced from the anomaly of electron and from the ratio $h/m_{\rm Cs}$ and $h/m_{\rm Rb}$. The value deduced of $h/m_{\rm Rb}$ correspond to our 2005 measurement \cite{{Clade1},{Clade2}}. On this figure, we have taken into account the recent correction of the QED calculation of the electron anomaly \cite{Gabrielse2007}.}
\label{fig:CODATA02EtNous}       
\end{figure}
These different measurements are shown on the figure \ref{fig:CODATA02EtNous}. The most precise determinations of $\alpha$ are deduced from the measurements of the electron anomaly $a_e$ made in the eighty's at the University of Washington \cite{VanDick} and, very recently, at Harvard \cite{{Gabrielse},{Odom}}. This last experiment and an impressive improvement of the QED calculation \cite{Kino} have lead to a new determination of $\alpha$ with a relative uncertainty of 0.70 ppb. Nevertheless this last determination of $\alpha$ has been corrected in 2007 because of a QED calculation error \cite{Gabrielse2007}. This shows the need of other determinations of $\alpha$ independent of the QED calculations, such as the values deduced from the the measurements of $h/m_{\rm Cs}$ \cite{Wicht} and $h/m_{\rm Rb}$  ($m_{\rm Cs}$ and $m_{\rm Rb}$ are the mass of Cesium and Rubidium atoms) which are also indicated on the figure \ref{fig:CODATA02EtNous}.

In this paper, we report the determination of $\alpha$ deduced from the ratio $h/m_{\rm Rb}$
\cite{{Battesti},{Clade1},{Clade2}}. Indeed, the fine structure constant can be
related to the ratio $h/m_X$ \cite{Taylor} by:
 \begin{equation}
\alpha^2=\frac{2R_\infty}{c}\frac{A_r(X)}{A_r(e)}\frac{h}{m_X}
\label{alpha-h/m}
\end{equation}
where $R_\infty$ is the Rydberg constant, $A_r(e)$ is the relative atomic mass of the electron and $A_r(X)$ the relative mass of the particle $X$ with mass $m_X$. These factors are known with a relative uncertainty of $7\times 10^{-12}$ for $R_\infty$ \cite{{Udem},{Schwob}}, $4.4\times 10^{-10}$ for $A_r(e)$ \cite{Beir} and less than $2.0\times 10^{-10}$ for $A_r(\rm Cs)$ and $A_r(\rm Rb)$ \cite{Bradley}. Hence, the factor limiting the accuracy of $\alpha$ is the ratio $h/m_X$.
\section{Principle of the experiment}
\label{sec:1}
The principle of the experiment is the measurement of the recoil velocity $v_r$ of a Rubidium atom absorbing or emitting a photon ($\mathbf{v_r}=\hbar\mathbf{k}/m$ , where $\mathbf{k}$ is the wave vector of the photon absorbed by the atom of mass $m$). Such a measurement yields to a determination of $h/m$ which can be used to obtain a value of the fine structure constant from equation \ref{alpha-h/m}.

\begin{figure}
\centering
\includegraphics[width=8cm]{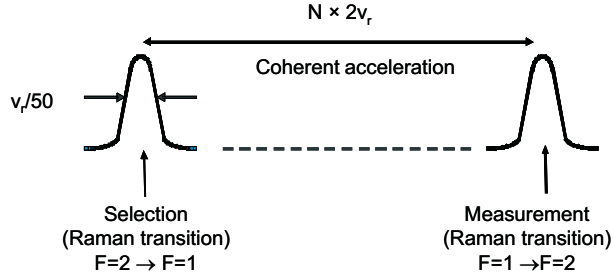}
\caption{We select a narrow velocity class by a first Raman transition between the $F=2$ and $F=1$ hyperfine sublevels of the ground state. Then the atoms are accelerated by means of Bloch oscillations, and the final velocity of the atoms is measured with a second Raman transition.}
\label{fig:Principe}       
\end{figure}

The experiment develops in three steps (see figure \ref{fig:Principe}). Firstly, we select from a cold atomic cloud a bunch of atoms with a very narrow velocity distribution. This selection is performed by a Doppler velocity sensitive counter-propagating Raman transition (a $\pi$-pulse transfers the atoms from the $F=2$ to the $F=1$ hyperfine level). Secondly, we transfer to these selected atoms as many recoils as possible by means of Bloch oscillations as explained latter. Finally, we measure the final velocity of the atoms by a second Raman transition which transfers the atoms from the $F=1$ to the $F=2$ hyperfine level.

\begin{figure}
\centering
\includegraphics[width=8cm]{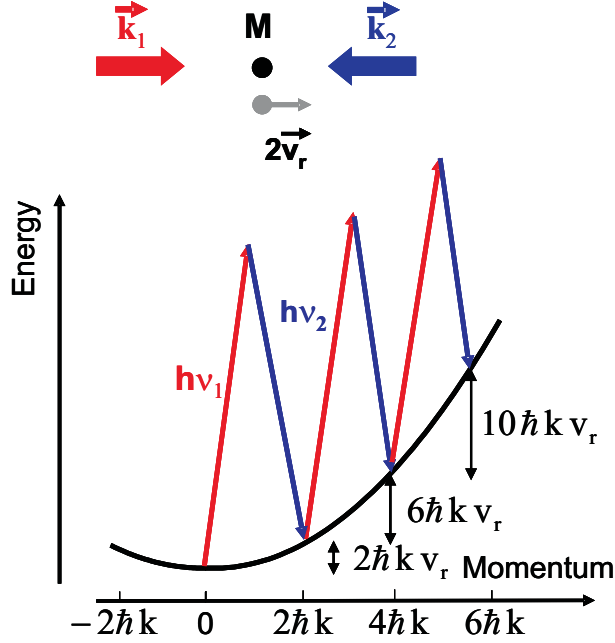}
\caption{Acceleration of cold atoms with a frequency chirped
standing wave. The variation of energy versus momentum in the
laboratory frame is given by a parabola. The energy of the atoms
increases by the quantity $4(2j+1)E_{r}$ at each cycle.}
\label{fig:Parabole}       
\end{figure}

The Bloch oscillations have been first observed in atomic physics by C. Salomon and coworkers \cite{{Dahan},{Peik}}. In a simple way, Bloch oscillations can be seen as Raman transitions where the atom begins and ends in the same energy level, so that its internal state ($F=1$) is unchanged while its velocity has increased by $2v_r$ per Bloch oscillation (see figure \ref{fig:Parabole}). Bloch oscillations are produced in one dimension optical lattice which is accelerated by linearly sweeping the relative frequencies of two counter propagating laser beams. The frequency difference $\Delta\nu$ is increased so that, because of the Doppler effect, the beams are always resonant with the same atoms ($\Delta\nu=4(2j+1)E_{r}/h$, $j=0,1,2,3..$ where $E_{r}/h$ is the recoil energy in frequency units and $j$ the number of transitions). This leads to a succession of rapid adiabatic passages between momentum states differing by $2h\nu/c$. In the solid-state physics approach, this phenomena is known as Bloch oscillations in the fundamental energy band of a periodical optical potential. The atoms are subject to a constant inertial force obtained by the introduction of the tunable frequency difference $\Delta\nu$ between the two waves that create the
optical potential \cite{Dahan}.

The accuracy of our measurement of the recoil velocity relies in the number of recoils ($2N$) that we are able to transfer to the atoms. Indeed, if we measure the final velocity with an accuracy of $\sigma_v$, the accuracy on the recoil velocity measurement $\sigma_{vr}$ is:
 \begin{equation}
\sigma_{vr}=\frac{\sigma_v}{2N}
\label{incertitude}
\end{equation}

\begin{figure}
\centering
\includegraphics[width=8cm]{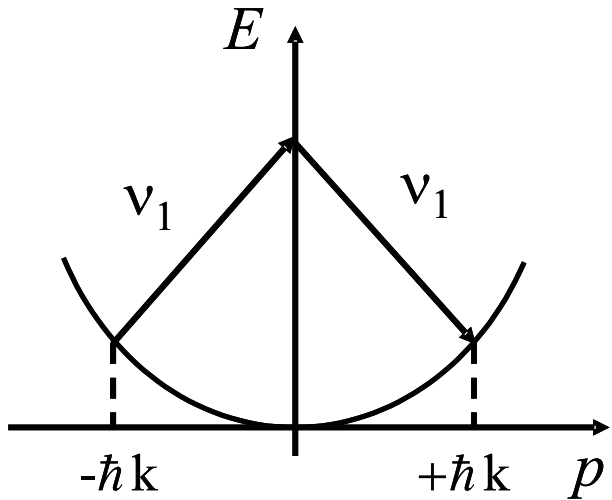}
\caption{Bloch oscillations due to the gravity in a vertical standing wave. }
\label{fig:BlochVertical}       
\end{figure}
Because of the acceleration due to gravity, Bloch oscillations can also be observed when an atom is placed in a vertical standing wave (see figure \ref{fig:BlochVertical}). The atom, initially at rest, starts to
fall because of gravity. When its momentum has reached the value $-\hbar k$, it absorbs a photon from the up-propagating wave and emits another one in the down-propagating wave. At the end of this $\Lambda$ transition its momentum is equal to $+\hbar k$. The atomic momentum varies between $+\hbar k$ and $-\hbar k$. The time required for this oscillation is equal to $T=2h\nu/cMg$ where $g$ is the acceleration due to gravity. In the case of rubidium atom, the oscillation frequency is about 830~Hz. This effect has been observed and briefly described in a previous paper \cite{stat}. It has also been reported by other groups \cite{{Inguscio},{Tino},{Naegerl}}.

\section{Experimental setup}
\label{sec:2}
\begin{figure}
\centering
\includegraphics[width=8cm]{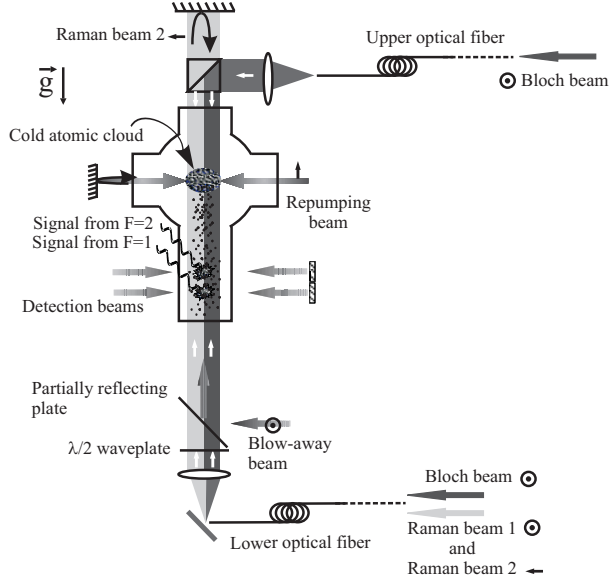}
\caption{Scheme of the experimental setup: the cold atomic cloud is
produced in a MOT (the cooling laser beams are not shown). The Raman
and the Bloch beams are vertical. The Raman beams and
the upward Bloch beam are injected into the same optical fiber. The
``blow-away'' beam is tuned to the one photon transition and allows us
to clear the atoms remaining in $F=2$ after the selection step. The
populations in the hyperfine levels $F=1$ and $F=2$ are detected  by
fluorescence 15~cm below the MOT using a time of flight
technique.} \label{fig:beams-setup}
\end{figure}

Our experimental setup has been previously described in detail \cite{Clade2}. It is shown on figure \ref{fig:beams-setup}. Briefly, $^{87}\rm Rb$ atoms are captured, from a background vapor, in a $\sigma^+-\sigma^-$ configuration magneto-optical trap (MOT). The trapping magnetic field is switched off and the atoms are cooled to about $3~\mu K$ in an optical molasses. After the cooling process, we apply a bias field of $\sim100~\mathrm{mG}$. The atoms are then optically pumped into the $F=2, m_F=0$ ground state. The determination of the velocity distribution is performed using a $\pi-\pi$ pulses sequence of two vertical counter-propagating laser beams (Raman beams) \cite{SENV}: the first pulse with a fixed frequency $\delta_{\rm sel}$, transfers atoms from $5 S_{1/2}$, $\left|F=2, m_F = 0\right>$ state to $5 S_{1/2}$,
$\left|F=1,m_F = 0\right>$ state, into a velocity class of about $v_r/15$ centered around $(\lambda \delta_{\rm sel}/ {2}) - v_r$ where $\lambda$ is the laser wavelength. To push away the atoms remaining in the ground state $F=2$, we apply after the first $\pi$-pulse, a resonant laser beam with the $5S_{1/2}~(F=2)$ to $5P_{1/2}~(F=3)$ cycling transition. Atoms in the state $F=1$ make $N$ Bloch oscillations in a vertical accelerated
optical lattice. We then perform the final velocity measurement using the second Raman $\pi$-pulse, whose frequency is $\delta_{\rm meas}$. The populations ($F=1$ and $F=2$) are measured separately by using the one-dimensional time of flight technique. To plot the final velocity distribution we repeat this procedure by scanning the Raman beam frequency $\delta_{\rm meas}$ of the second pulse.

The Raman beams are produced by two stabilized diode lasers. Their beat frequency is precisely controlled by a frequency chain allowing to easily switch from the selection frequency ($\delta_{\rm sel}$) to the measurement frequency ($\delta_{\rm meas}$). One of the lasers is stabilized on a highly stable Fabry-Perot
cavity and its frequency is continuously measured by counting the beatnote with our two-photon Rb standard \cite{DeuxPhotonRb}. The two beams have linear orthogonal polarizations and are coupled into the same
polarization maintaining optical fiber. The pair of Raman beams is sent through the vacuum cell. The counter propagating configuration is achieved using a polarizing beam-splitter cube and an horizontal retroreflecting mirror placed above the exit window of the cell.

The standing wave used to create the 1-D optical lattice is generated by a Ti:Saphire laser, whose frequency is stabilized on the same highly stable Fabry-Perot cavity. This laser beam is split in two parts. To perform the timing sequence, each one passes through an acousto-optic modulator to control its intensity and frequency. The optical lattice is blue detuned by $\sim 40~\mathrm{GHz}$ from the one photon transition. It is adiabatically raised (500~$\mu$s) in order to load all the atoms into the first Bloch band. To perform
the coherent acceleration, the frequency difference of the two laser beams generating the optical lattice is swept linearly. Then, the lattice intensity is adiabatically lowered (500~$\mu$s) to bring atoms back in a well defined momentum state. The optical potential depth is $70~E_r$. With these lattice parameters, the spontaneous emission is negligible. For an acceleration of about $2000~\rm m\rm s^{-2}$ we transfer about 900 recoil momenta in 3~ms with an efficiency greater than $99.97\%$ per recoil. To avoid atoms from
reaching the upper windows of the vacuum chamber, we use a double acceleration scheme: instead of selecting atoms at rest, we first accelerate them using Bloch oscillations and then we make the three steps sequence: selection-acceleration-measurement. This way, the atomic velocity at the measurement step is close to zero. In the vertical direction, an accurate determination of the recoil velocity would require an accurate measurement of the gravity $g$. In order to get rid of gravity, we make a differential measurement by accelerating the atoms in opposite directions (up and down trajectories) keeping the same delay between the selection and the measurement $\pi$-pulses. The ratio $\hbar/m$ can then be deduced
from the formula:
\begin{equation}
\frac{\hbar}{m}= \frac{(\delta_{\rm sel}-\delta_{\rm meas})^{\rm up} -
(\delta_{\rm sel}-\delta_{\rm meas})^{\rm down}}{2(N^{\rm up}+N^{\rm down})k_B
(k_1+k_2)} \label{eq:mynum}
\end{equation}
where $(\delta_{\rm meas}-\delta_{\rm sel})^{\rm up/down}$ corresponds respectively to the center of the final velocity distribution for the up and the down trajectories, $N^{\rm up/down}$ are the number of Bloch oscillations in both opposite directions, $k_B$ is the Bloch wave vector and $k_1$ and $k_2$ are the wave vectors of the Raman beams.

Moreover, the contribution of some systematic effects (energy level shifts) to $\delta_{\rm sel}$ and $\delta_{\rm meas}$ is inverted when the direction of the Raman beams are exchanged. To improve the experimental protocol, for each up or down trajectory, the Raman beams directions are reversed and we record two velocity spectra. Finally, each determination of  $h/m_{\rm Rb}$  and $\alpha$ is obtained from 4 velocity spectra.

\section{Results in the $\pi$-$\pi$ configuration}
\label{sec:3}

\def\unite#1{{\mathrm{~#1}}}
\begin{figure}
  \begin{minipage}{.5\textwidth}
   \centering
    $\delta = 13\,614\,059,9\pm 1,6\unite{Hz}$ \\
  \includegraphics[width=0.95\textwidth]{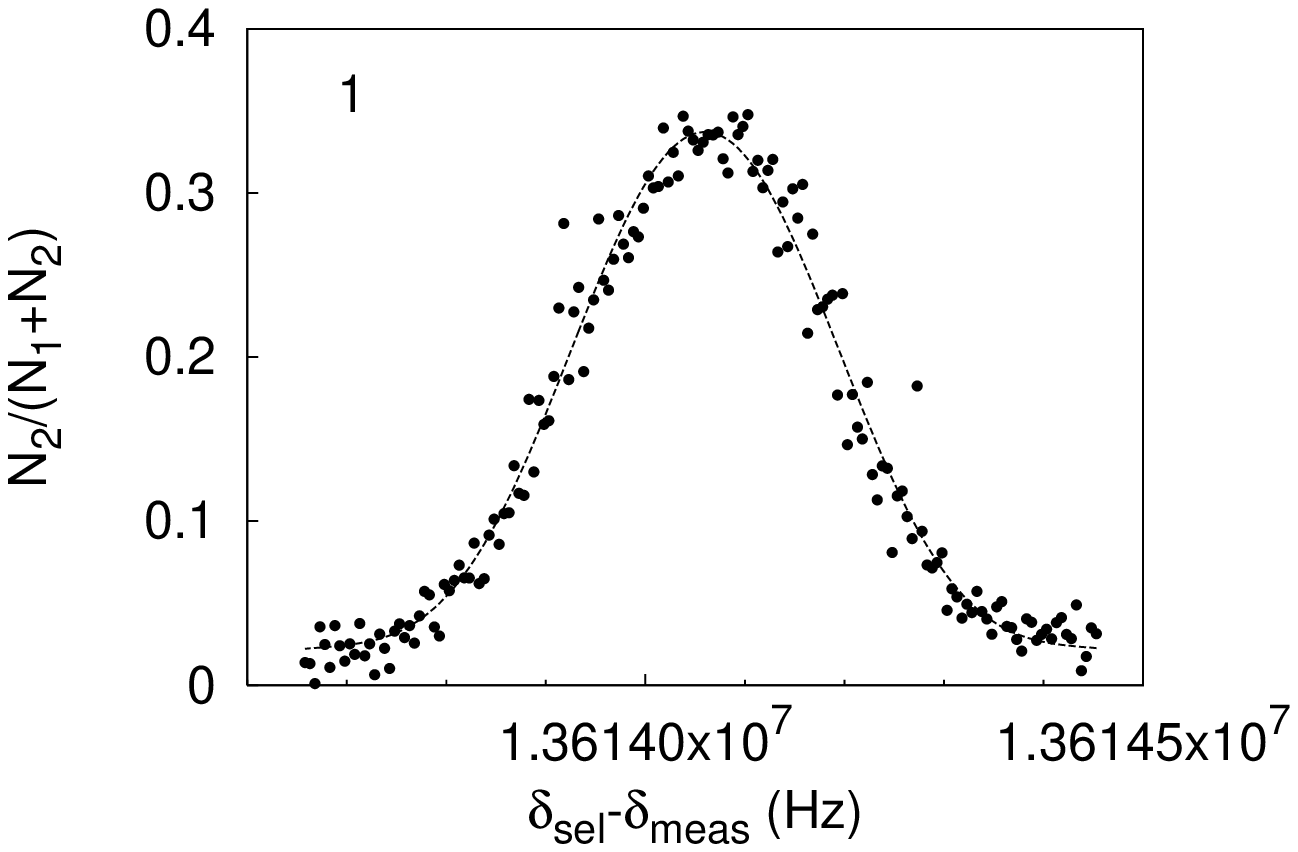}\\
  \end{minipage}
  \begin{minipage}{.5\textwidth}
   \centering
    $\delta = -13\,311\,534,1\pm 1,6\unite{Hz}$ \\
  \includegraphics[width=0.95\textwidth]{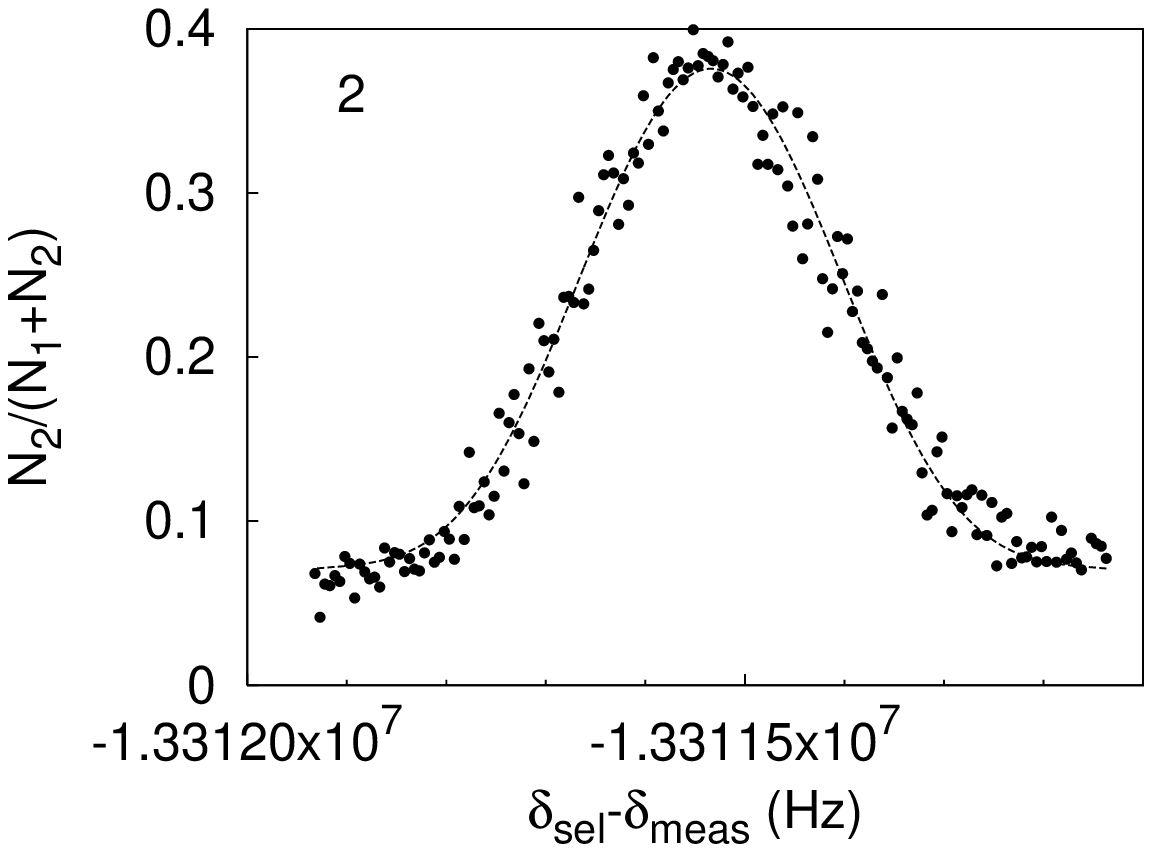}\\
  \end{minipage} \\
  \medskip \\
  \begin{minipage}{.5\textwidth}
     \centering
    $\delta = 13\,614\,031,2\pm 1,8\unite{Hz}$ \\
  \includegraphics[width=0.95\textwidth]{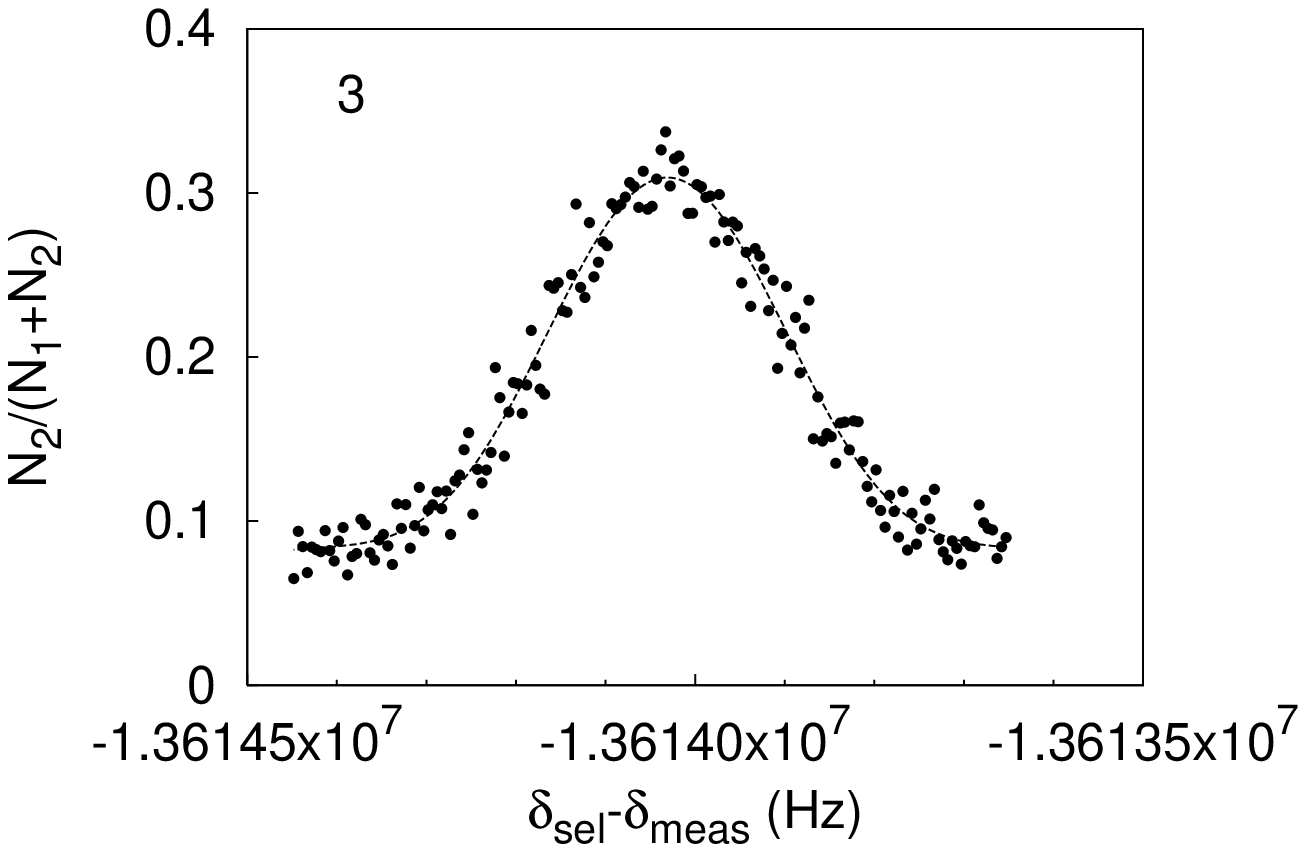}\\
  \end{minipage}
  \begin{minipage}{.5\textwidth}
   \centering
   $\delta = -13\,311\,588,1\pm 1,7\unite{Hz}$ \\
  \includegraphics[width=0.95\textwidth]{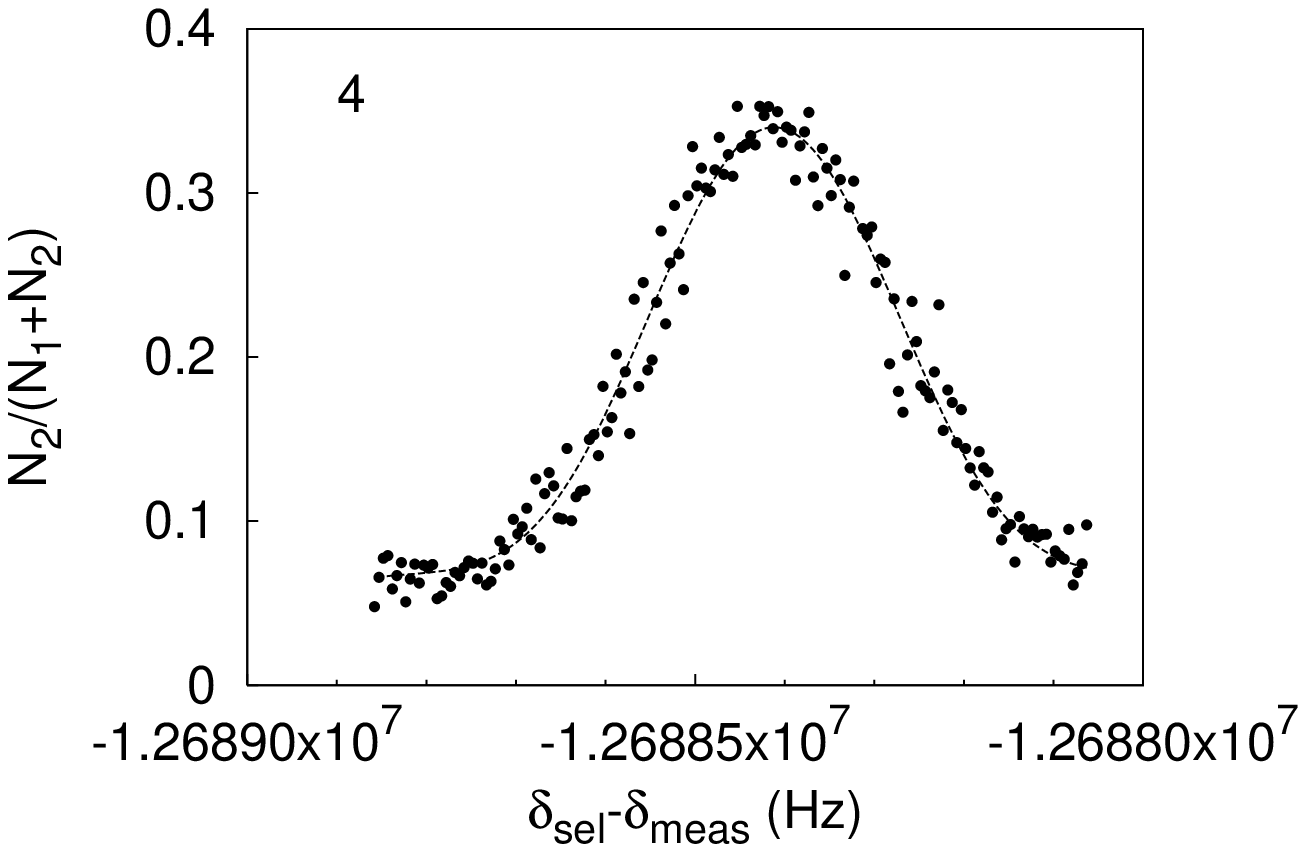}\\
  \end{minipage}
\caption{Sequence of four spectra used for each determination of $\alpha$. Here $N_1$ and $N_2$ are respectively the number of atoms in $F=1$ and $F=2$ after the acceleration process. The spectra are obtained by exchanging the Raman beams direction and performing the Bloch acceleration upwards or downwards. The relative uncertainty for each spectrum is about $1.7$ Hz. From these four spectra, $h/m_{\rm Rb}$ can be determined with an uncertainty of $6.6\times 10^{-8}$.}
\label{fig:quatre-spectre1}
\end{figure}

We present in this section the results obtained in 2005 following the protocol described above. The  determinations of $ h/m_{\rm Rb}$ and $\alpha$ have been derived from 72 experimental data point taken during four days. Each value of $h/m_{\rm Rb}$ is obtained from four spectra as detailed in the previous section (see figure \ref{fig:quatre-spectre1}). The uncertainty in the determination of the central frequency of each spectrum is about 1.7 Hz ($\simeq10^{-4}v_r$). In these measurements, the number of Bloch oscillations were $N^{\rm up}=430$ and $N^{\rm down}=460$. Then, the effective recoil number is $2(N^{\rm up}+N^{\rm down})=1780$.

\begin{figure}
\centering
\includegraphics[width=8cm]{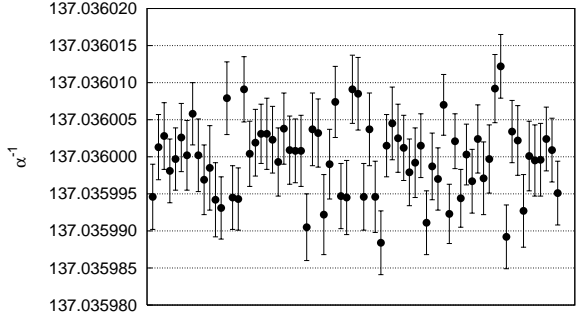}
\caption{Chronologically, our 72 determinations of the fine structure constant made in 2005 with the $\pi$-$\pi$ configuration.}
\label{fig:Alpha2005}
\end{figure}

We present on figure \ref{fig:Alpha2005} the set of 72 determinations of the fine structure constant $\alpha$. In each one of them, we have transferred to the atoms up to 460 Bloch oscillations, with an efficiency of 99.95$\%$ per oscillation. Each determination is obtained after 20 minutes of integration time. The corresponding relative uncertainty in $h/m_{\rm Rb}$ is around $6.6\times 10^{-8}$ and hence $\alpha$ is deduced with a relative uncertainty of $3.3\times 10^{-8}$. The dispersion of these $n=72$ measurements is $\chi^2/(n-1)=1.3$ and the resulting statistical relative
uncertainty on $h/m_{\rm Rb}$ is 8.8 ppb. This corresponds to a relative statistical uncertainty on $\alpha$ of 4.4 ppb.

All systematic effects affecting the experimental measurement have been analyzed in detail in the reference \cite{Clade2}. They are recall in table \ref{tab:1}. Here we discuss only the two main corrections which are due to the geometry of the laser beams and to the second order Zeeman effect. As the experimental laser beams are not plane waves, we have to consider the phase gradient in equation \ref{eq:mynum} instead of wave vector $k$. For a Gaussian beam the effective wave vector is given by \cite{Kogelnik}:

\begin{equation}\label{phase}
    k_z^\mathrm{eff} = \frac{\mathrm{d}\phi}{\mathrm{d}z} = k -
    \frac2{kw(z)^2}\left(1-\frac{r^2}{w(z)^2}\left(1-(z/z_R)^2\right)\right)
\end{equation}
\\
where $z_R = \pi w_0^2/\lambda$ is the Rayleigh length and $w(z)^2=w_0^2(1+(z/z_R)^2)$. Equation (\ref{phase}) gives the effective wave vector as a function of two parameters: $w(z)$ the waist of the beam at the measurement point and $z/z_R$. To evaluate these two parameters, we have used a Shack-Hartmann wave front analyzer (HASO 64, Imagine Optic) which measures the wavefront curvature radius $R(z)$ and the waist $w(z)$ at a given position. From these measurements, we have obtained a correction of 16.4 $\pm$ 8.0 ppb on the determination of $h/m_{\rm Rb}$.

As explained in section \ref{sec:2}, the effect of parasitic level shifts is eliminated by inverting the direction of the Raman beams. Indeed, we have:

\begin{equation}\label{levelshift}
    \delta_{\rm meas}-\delta_{\rm sel}=2Nv_r(k_1+k_2)+\Delta_d(z,t)/h
\end{equation}
\\
where $\Delta_d(z,t)$ is the differential level shift between the measurement and the selection, which can depend on the time and the positions. From equation \ref{levelshift}, we see that the cancelation of the Zeeman level shifts assumes that between two consecutive measurements the magnetic field  and the atomic position are the same. The temporal sequence being the same for the two directions of propagation, there is no reason for a systematic effect due to a temporal variation of the magnetic field. However, the position of atoms is not exactly the same because the directions of the recoils given at the first Raman transition are opposite. For the timing used in our experiment, they differ by about  $\delta_z=300~\mu\mathrm m$. We have precisely measured the spatial magnetic field variations using copropagating Raman transitions. This leads to a differential level shift of 0.3 $\pm$ 0.1 Hz and a correction of -13.2 $\pm$ 4.0 ppb on the determination of $h/m_{\rm Rb}$.

\begin{table}
\caption{Error budget on the determination of $h/m_{\rm Rb}$ (correction and relative uncertainty in ppb).}
\label{tab:1}       
\begin{tabular}{lll}
\hline\noalign{\smallskip}
Source & Correction & Relative uncertainty (ppb)  \\
\noalign{\smallskip}\hline\noalign{\smallskip}
Laser frequencies& &1.6\\
Beams alignment&4& 4\\
Wavefront curvature and Gouy phase&16.4 & 8\\
2nd order Zeeman effect&-13.2 & 4 \\
Quadratic magnetic force&2.6 & 0.8\\
Gravity gradient&0.36& $0.04$ \\
Light shift (one photon transition)& & 0.4\\
Light shift (two photon transition)&1.0 & 0.4 \\
Light shift (Bloch oscillation)&-0.92 & 0.4 \\
Index of refraction atomic cloud& &0.6 \\
Index of refraction background vapor&0.75 & 0.6 \\
\hline\noalign{\smallskip}
Global systematic effects&10.98 & 10.0\\
\noalign{\smallskip}\hline
\end{tabular}
\end{table}

Finally, the total correction due to the systematic effects is 10.98 $\pm$ 10.0 ppb on the determination of $h/m_{\rm Rb}$. With this correction, we obtain for $h/m_{\rm Rb}$ and $\alpha$:

\begin{equation}
\frac{h}{m_{\rm Rb}}= 4.591\,359\,29\,(6)\times10^{-9}\quad [1.3\times
10^{-8}]~~~\rm m^2\rm s^{-1}
\end{equation}

\begin{equation}
    \alpha^{-1}= 137.035\,998\,84\,(91) \quad [6,7\times 10^{-9}]
\end{equation}

This value of the fine structure constant is labeled $h/m(\rm Rb)$ on the figure \ref{fig:CODATA02EtNous}. It is in agreement with the values deduced from the measurement of $h/m_{\rm Cs}$ and from the electron anomaly.

\section{Measurement of the fine structure constant by atomic interferometry}
\label{sec:4}

\begin{figure}
\centering
\includegraphics[width=8cm]{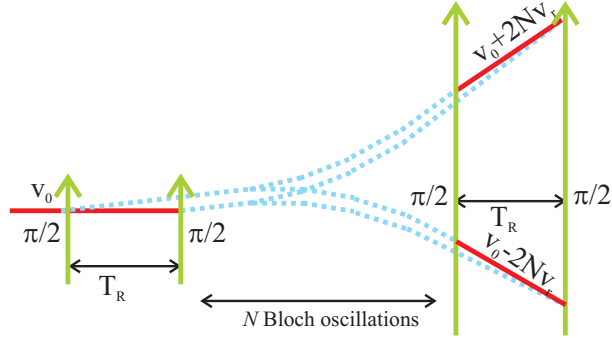}
\caption{Scheme of the interferometers used for the measurement of $h/m_{\rm Rb}$. The first pair of $\pi/2$ pulses selects a comb of velocities which is measured by the second pair of $\pi/2$ pulses. Between these two pairs of pulses, the atoms are accelerated upwards or downwards. The solid line corresponds to the atom in the $F=2$ state, and the dashed line to the $F=1$ state.}
\label{fig:Interferometre}
\end{figure}

We present in this section the measurement of $\alpha$ using atomic interferometry. In 2007, we have modified the experimental scheme to take advantage of Ramsey spectroscopy. For the selection, we use a pair of $\pi/2$ pulses which select a comb of velocities. Then, the atoms are accelerated upwards or downwards as precedently by using Bloch oscillations. Finally the measurement is made by a second pair of $\pi/2$ pulses. The scheme of this interferometer is shown on figure \ref{fig:Interferometre}. The frequency resolution is now determined by the time within each pair of pulses $T_R$ while the duration of each $\pi/2$ pulse determines the spectral width of the pulses and the number of atoms which contribute to the signal. This interferometer is similar to the one of the reference \cite{Wicht}, except that effective Ramsey k-wavevectors point in the same direction. Consequently, this interferometer is not sensitive to the recoil energy, but only to the velocity variation due to the Bloch oscillations which take place between the two sets of $\pi/2$ pulses.

\begin{figure}
\begin{minipage}{.5\textwidth}
   \centering
    $\delta = -12\,380\,012.1\pm 1.2\unite{Hz}$ \\
  \includegraphics[width=0.95\textwidth]{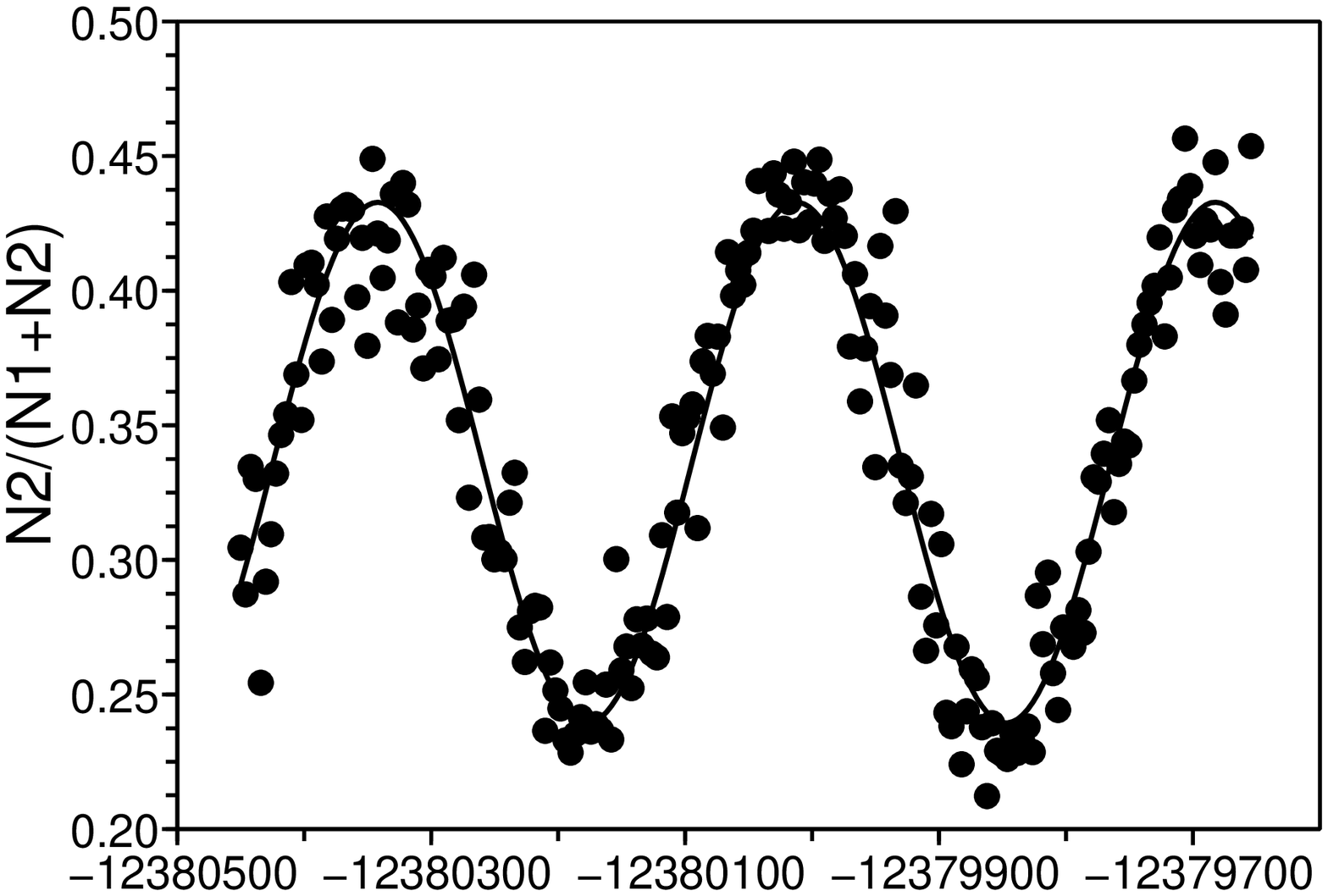}
  \end{minipage}
  \begin{minipage}{.5\textwidth}
   \centering
    $\delta = 17\,815\,192.1\pm 1.4\unite{Hz}$ \\
  \includegraphics[width=0.95\textwidth]{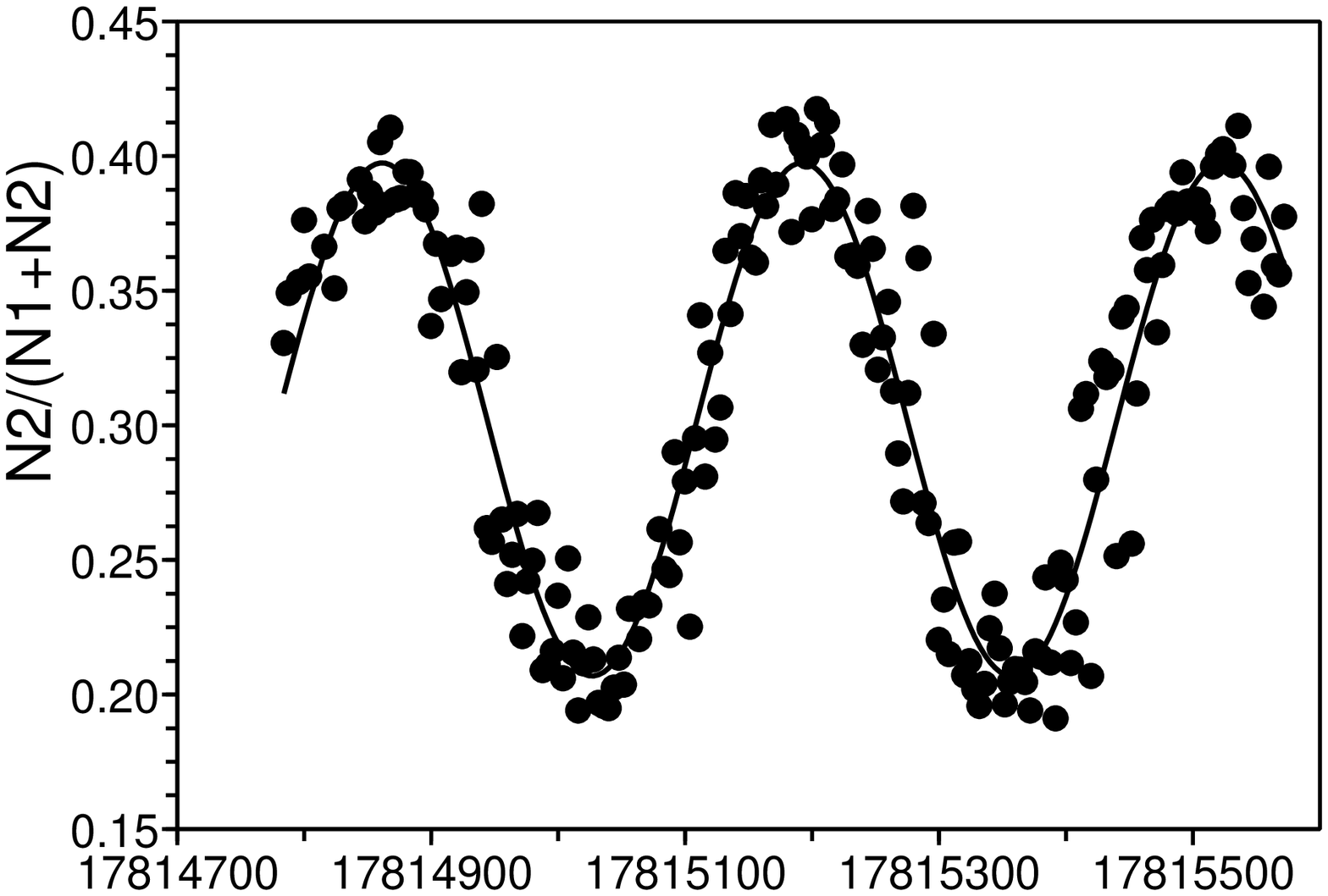}\\
  \end{minipage} \\
  \medskip \\
  \begin{minipage}{.5\textwidth}
     \centering
    $\delta = 12\,410\,319.5\pm 1.3\unite{Hz}$ \\
  \includegraphics[width=0.95\textwidth]{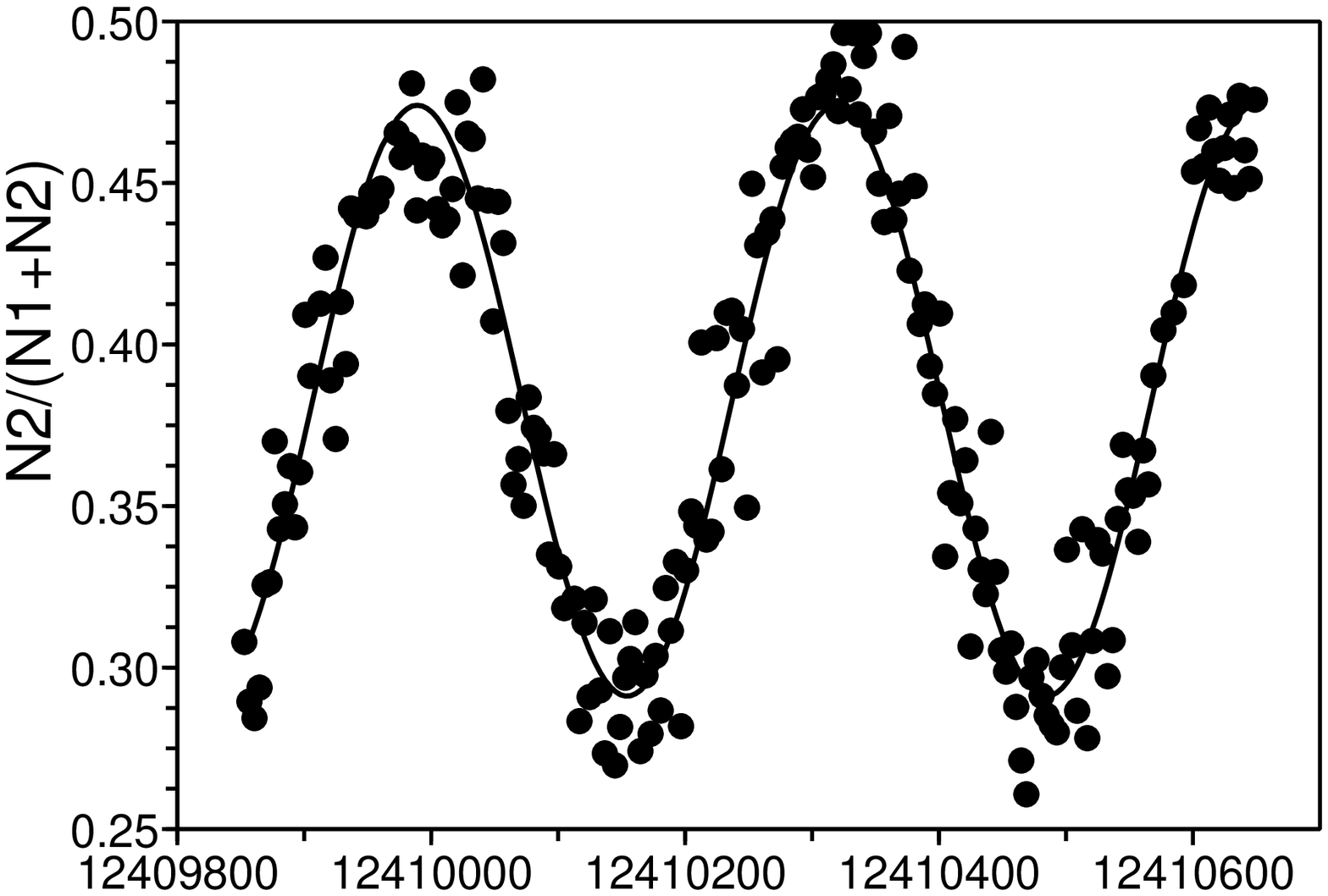}\\
  \end{minipage}
  \begin{minipage}{.5\textwidth}
   \centering
   $\delta = -17\,784\,943.7\pm 1.7\unite{Hz}$ \\
  \includegraphics[width=0.95\textwidth]{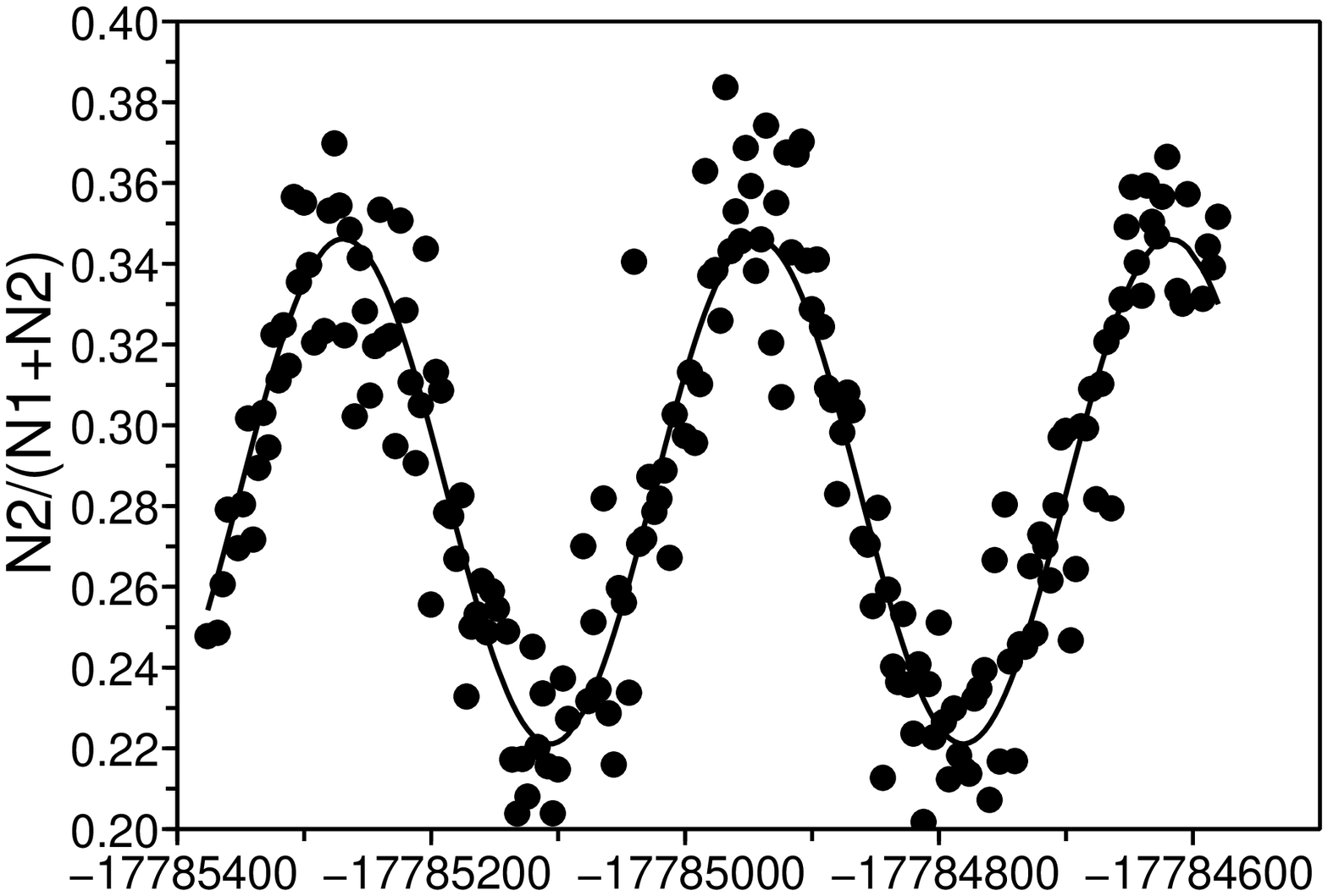}\\
  \end{minipage}
\caption{Spectra obtained by atomic interferometry which show the proportion of atoms in the $F=2$ level ($N_2/(N_1+N_2)$) in function of the frequency difference $\delta_{\rm meas}-\delta_{\rm sel}$ in Hz. The two spectra on the left corresponds to the upwards acceleration (400 Bloch oscillations) and the two spectra on the right to the downwards acceleration (600 Bloch oscillations).  }
\label{fig:quatre-spectre2}
\end{figure}

As in the precedent experiment, a value of $h/m_{\rm Rb}$ is deduced from four spectra obtained with the upwards or downwards acceleration and by exchanging the direction of the Raman beams. An example is shown on the figure \ref{fig:quatre-spectre2}. In this case, the total number of Bloch oscillations is $N^{\rm up}+N^{\rm down}=1000$, corresponding to 2000 recoil velocities between the up and down trajectories. The duration of each $\pi/2$ pulse is 400 $\mu$s and the time $T_R$ is 2.6 ms (the total time of a pair of $\pi/2$ pulses is 3.4 ms). For these experiments, the blue detuning of the Raman lasers is 310 GHz.

We have recorded numerous such spectra. For these measurements, the total number of Bloch oscillations $N^{\rm up}+N^{\rm down}$ varies from 200 to 1100. We are analyzing these data to confirm the preliminary result given at the Atomic Clocks and Fundamental Constants conference. The final result will be published subsequently.

\begin{figure}
\centering
\includegraphics[width=12cm]{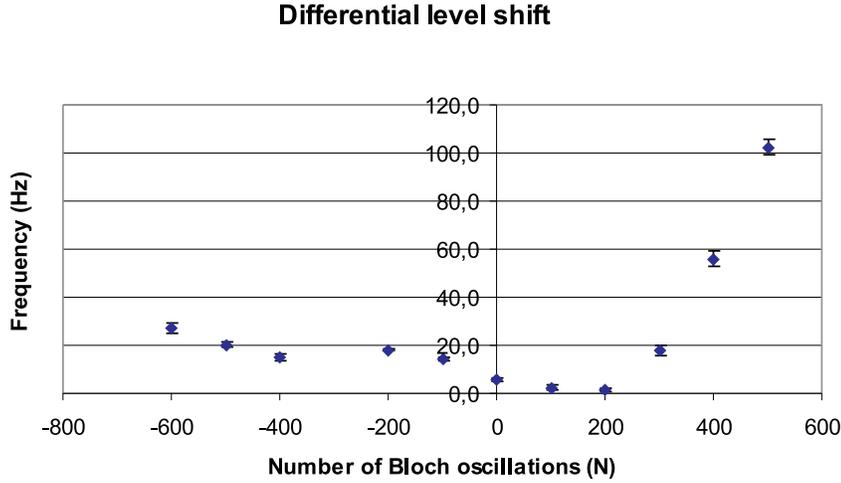}
\caption{Variation of the level shift $\Delta_d(z,t)$ with the number of Bloch oscillations. As the position of the atoms during the selection and the measurement are directly linked to the number of Bloch oscillations, we can deduce the variation of this shift with the position.}
\label{fig:Zeeman}
\end{figure}

For these measurements, the corrections due to the systematic effects will be very similar to the ones of table \ref{tab:1}. Nevertheless, we have developed a new method to obtain a map of the level shifts due mainly to the second-order Zeeman effect. Indeed, we can extract this information from the the four spectra presented on the figure \ref{fig:quatre-spectre2}. When we exchange the direction of the Raman beams, we obtain two equations similar to equation \ref{levelshift} from where we can deduce the value of the differential level shift $\Delta_d(z,t)$. More precisely, we have:

\begin{equation}\label{levelshift1}
    \Delta_d(z,t)=\Delta(z_{\rm meas},t)-\Delta(z_{\rm sel},t)
\end{equation}
\\
where $\Delta(z,t)$ is the level shift which depends on the position $z$, $z_{\rm meas}$ and $z_{\rm sel}$ the positions during the measurement and the selection. As we can change these positions by varying the number of Bloch oscillations, we are able to deduce precisely the variation of this shift with the position. This is illustrated on figure \ref{fig:Zeeman} which shows the variation of $\Delta_d(z,t)$ with the number of Bloch oscillations (for these measurements, the number of Bloch oscillations during the acceleration and the deceleration are equal). With this method, we have reduced the uncertainty due to this correction by a factor two.

\section{Conclusion}
\label{sec:5}

We have described two methods to measure the ratio $h/m_{\rm Rb}$. Depending on the Raman pulses arrangement ([$\pi/2$-$\pi/2$]-[$\pi/2$-$\pi/2$] or [$\pi$]-[$\pi$]), our experiment can run as an atom interferometer or not.  The comparison of our result with the value extracted from the electron magnetic anomaly experiment \cite{{Gabrielse},{Odom}} is either a strong test of QED calculations or, assuming these calculations exact, it gives a limit to test a possible internal structure of the electron.

To improve this test, our goal is now to reduce the uncertainty. We are building a new experimental setup with a larger vacuum chamber. Indeed, in our current setup, we cannot increase the number of Bloch oscillations because of the displacement of the atoms in the vacuum cell. With the new cell, we plan to multiply the number of Bloch oscillations by a factor three. We will use also a magnetic shield to reduce the parasitic magnetic field and a double cell with a 2D-MOT which loads a 3D-MOT in order to reduce the density of the background vapor in the interaction area. With these different improvements, we plan to reduce the uncertainty at the level of one ppb to obtain an unprecedent test of the QED calculations.

This experiment is supported in part by the Laboratoire National de M\'etrologie et d'Essais (Ex. Bureau
National de M\'etrologie) (contract 033006), by IFRAF (Institut Francilien de Recherches sur les Atomes
Froids) and by ANR (Agence Nationale de la Recherche, FISCOM project).
%

%
%

\end{document}